# A two-dimensional spin field-effect transistor

Wenjing Yan[1,*], Oihana Txoperena[1,*], Roger Llopis[1], Hanan Dery[2,3], Luis E.Hueso[1,4,†], and Fèlix Casanova[1,4,†]

[1]CIC nanoGUNE, 20018 Donostia-San Sebastian, Basque Country, Spain.

[2]Department of Electrical and Computer Engineering, University of Rochester, Rochester, New York, 14627, USA

[3]Department of Physics and Astronomy, University of Rochester, Rochester, New York, 14627, USA

[4] IKERBASQUE, Basque Foundation for Science, 48011 Bilbao, Basque Country, Spain.

*These authors contributed equally to this work

[†]E-mail: l.hueso@nanogune.eu; f.casanova@nanogune.eu

**The integration of the spin degree of freedom in charge-based electronic devices has revolutionised both sensing and memory capability in microelectronics[1]. Further development in spintronic devices requires electrical manipulation of spin current for logic operations. The approach followed so far, inspired by the seminal proposal of the Datta and Das spin modulator[2], has relied on the spin-orbit field as a medium for electrical control of the spin state [3-6]. However, the still standing challenge is to find a material whose spin-orbit-coupling (SOC) is weak enough to transport spins over long distances, while also being strong enough to allow their electrical manipulation. Here we demonstrate a radically different approach by engineering a heterostructure from atomically thin crystals, which are "glued" by weak van der Waals (vdW) forces[7] and which combine the superior spin transport properties of graphene[8-11] with the strong SOC of the semiconducting $MoS_2$[12-15]. The spin transport in the graphene channel is modulated between ON and OFF states by tuning the spin absorption into the $MoS_2$ layer with a gate electrode. Our demonstration of a spin field-effect transistor using**

**two-dimensional (2D) materials identifies a new route towards spin logic operations for beyond CMOS technology.**

Carbon-based materials with intrinsic weak spin-orbit coupling (SOC), such as organic semiconductors[16], carbon nanotubes[17] and graphene[18], have made a notable impact in spintronics. In particular, graphene has been proved to be ideal for long (in excess of several microns) distance spin transport[8-11]. However, due to its weak SOC, spin manipulation in this material has been mainly achieved by an external magnetic field through Hanle precession [8, 10-11]. Although various approaches have been taken to enhance the SOC of graphene, for example through proximity effect[19-20] or by atomic doping[21], a direct evidence on the modulation of spin transport by an electric field remains elusive.

Meanwhile, transition metal dichalcogenides (TMDs) have emerged to complement graphene due to their unique optical, spin and valley properties[14, 22]. Specifically, $MoS_2$, the best-known member of that class, has a crossover from an indirect to a direct-gap semiconductor when thinned down to a monolayer (ML)[23]. Its stronger SOC compared with that of graphene, arising from the *d*-orbitals of the transition metal atoms, offers new possibilities to employ the spin and valley degrees of freedom in TMDs[15]. The combination of graphene with $MoS_2$ in a vdW heterostructure allows us to engineer a new type of spin field-effect transistor (spin-FET), as shown in Fig. 1a. Spin current in the graphene section of the device is electrically injected from a ferromagnetic source terminal. The gate electrode controls how much of that spin current is absorbed in the intersecting $MoS_2$ layer (spin sink) prior to its arrival to the ferromagnetic drain terminal. Depending on the gate voltage, the detected spin signal is a binary ON/OFF spin current.

A scanning electron microscope image of the device is shown in Fig. 1b. Graphene flakes are exfoliated onto a highly-doped Si substrate covered by 300 nm of $SiO_2$. A monolayer graphene flake is identified according to its optical contrast[24] and, subsequently, a few-layer MoS$_2$ flake is transferred above it by all-dry viscoelastic stamping[25]. Several TiO$_2$/Co electrodes are patterned by electron-beam lithography and evaporated onto the graphene channel to create lateral spin valves (LSV) (see methods), which enable the injection and detection of pure spin currents in graphene in a non-local geometry[8,11]. The non-local resistance $R_{nl} = V_{nl}/I$, which depends on the relative orientation of the magnetisation of the injecting and the detecting Co electrodes, is measured while sweeping the magnetic field $B$ in-plane along the easy axis of the electrodes (see Fig. 1a for a sketch of the experimental geometry). Specifically, when the configuration of the magnetisations changes from parallel to antiparallel, $R_{nl}$ switches from high ($R_p$) to low ($R_{ap}$) values. The spin signal is proportional to the amount of spin current reaching the detector, measured by $\Delta R_{nl} = R_p - R_{ap}$ (Fig. 1c).

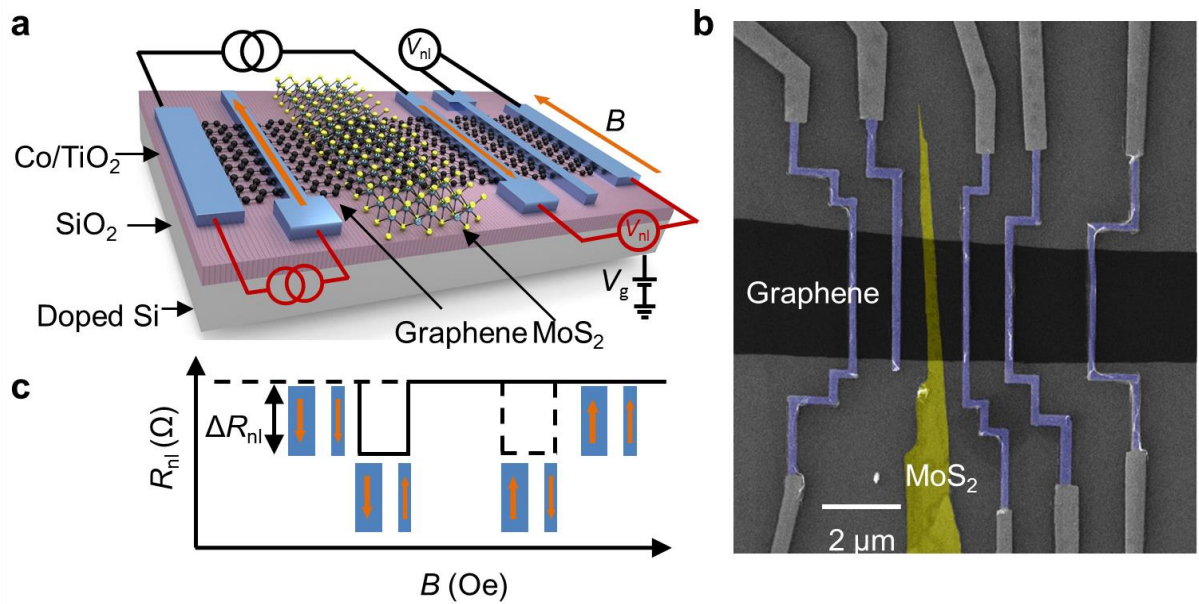

**Figure 1| Illustration of the experiment and scanning electron microscope (SEM) image of the devices**. **a**, Sketch of the spin field-effect transistor. For the non-local measurement, a DC current is injected into graphene from a ferromagnetic Co electrode across a TiO$_2$ barrier, while a non-local voltage ($V_{nl}$) is measured by a second Co electrode. The black- and red-coloured circuit diagrams represent the measurement configurations in the reference graphene LSV (without MoS$_2$ on top) and

the graphene/MoS$_2$ LSV (with MoS$_2$ intercepting the spin current path). In the later case, the spin current flowing in the graphene can be switched ON and OFF by modulating the conductivity of MoS$_2$ using an electric field across a SiO$_2$ dielectric (also shown in the diagram). **b**, False-coloured SEM image of the LSV devices. **c**, Sketch illustrating a typical non-local magnetoresistance measurement, where the non-local resistance $R_{nl}$ switches between $R_P$ and $R_{AP}$ for parallel and antiparallel magnetisation orientation of the Co electrodes. The spin signal is tagged as $\Delta R_{nl} = R_P - R_{AP}$.

We first study the spin transport in a graphene LSV without MoS$_2$ (reference LSV). Figure 2a shows the measured $R_{nl}$ as a function of $B$ for different gate voltages ($V_g$). Upon the application of $V_g$, the magnitude of the spin signal weakly varies, following the modification of the graphene sheet conductivity ($\sigma_{Gr}^{\blacksquare}$) with $V_g$, as can be observed in Fig. 2b. The correlation between $\Delta R_{nl}$ and $\sigma_{Gr}^{\blacksquare}$ is a signature of a transparent interface between the Co/TiO$_2$ electrodes and the graphene (~250 Ω), as it is well established in the literature[11].

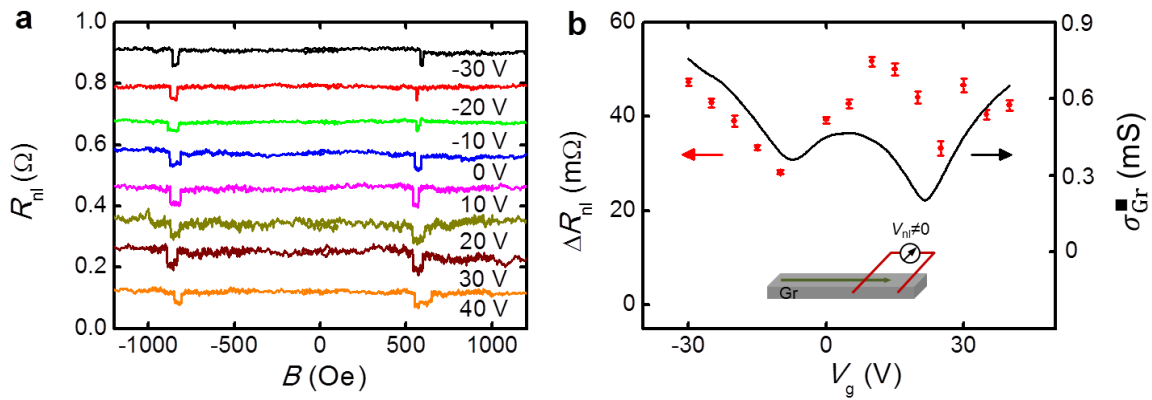

**Figure 2| Spin signal in a reference graphene lateral spin valve.** Measurements are done using the black-coloured circuit diagram in Fig.1a. **a**, Non-local resistance $R_{nl}$ as a function of the magnetic field $B$ measured at different $V_g$ at 50 K, using 10 µA current bias and for a centre-to-centre distance between ferromagnetic electrodes ($L$) of 1 µm. Individual sweeps are offset in $R_{nl}$ for clarity. **b**, Spin signal $\Delta R_{nl}$ measured at different $V_g$ (red squares). The black solid line shows the sheet conductivity of the graphene as a function of $V_g$. The inset indicates the spin current (green arrow) reaching the detector in the full range of $V_g$.

Next, we introduce the central results of our manuscript: the demonstration of a spin field-effect transistor in a graphene/MoS$_2$ LSV. Figure 3a shows $R_{nl}$ of this device while sweeping $B$ for different values of $V_g$, where a gradual decrease of the spin signal $\Delta R_{nl}$ with $V_g$ can be observed. This behaviour is clearly seen in Fig. 3b, where $\Delta R_{nl}$ is plotted as function of $V_g$, showing the decay of $\Delta R_{nl}$ towards zero at positive values of $V_g$, in contrast with the weakly varying signal measured in the reference LSV (see Fig. 2b). Figure 3b also plots the MoS$_2$ sheet conductivity ($\sigma_{MoS_2}^{\blacksquare}$) from a reference device revealing an opposite gate voltage dependence to that of $\Delta R_{nl}$. For large negative $V_g$, the semiconducting MoS$_2$ is in the low conductivity OFF state, and the measured $\Delta R_{nl}$ value is comparable to that of the reference LSV. This result is expected considering that the electrode spacing here ($L \approx 1.8$ μm) is slightly longer than in the reference LSV ($L \approx 1$ μm) (see Fig. 2b for comparison). Sweeping the gate voltage towards positive values brings the MoS$_2$ towards its high conductivity ON state, where $\sigma_{MoS_2}^{\blacksquare}$ increases by more than 6 orders of magnitude compared to the OFF state. Simultaneously, the spin current reaching the detector and the corresponding $\Delta R_{nl}$ gradually decrease towards zero (see Fig. 3b). The results are completely reproducible upon multiple gate voltage sweeps and temperature cycles, evidencing the robustness of the effect (see Supplementary Information). Similar results to those in Fig. 3b are also observed at temperatures up to 200 K (see Supplementary Information).

This control of the spin current constitutes the direct demonstration of a spin field-effect transistor. Following an analogy with conventional transistors, we can introduce the spin transconductance as a figure of merit in the spin field-effect transistor. It could be defined as the change in spin signal per gate voltage unit, yielding ~0.7 mΩ/V in the specific case here presented.

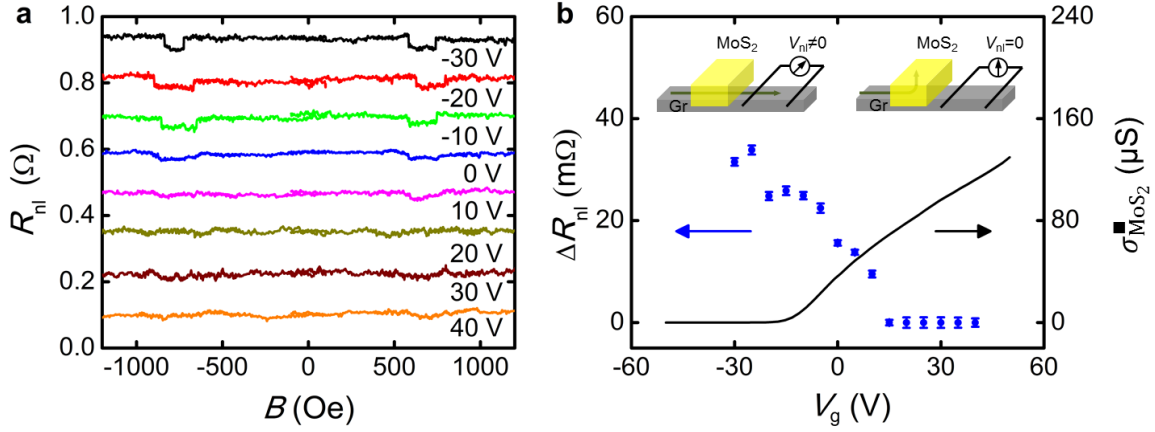

**Figure 3| Spin signal in a graphene/MoS$_2$ lateral spin valve.** Measurements are done using the red-coloured circuit diagram in Fig.1a. **a**, Non-local resistance $R_{nl}$ measured as a function of the magnetic field $B$ at different $V_g$ at 50 K using 10 µA current bias and for a centre-to-centre distance between ferromagnetic electrodes ($L$) of 1.8 µm. Individual sweeps are offset in $R_{nl}$ for clarity. **b**, Gate modulation of the spin signal $\Delta R_{nl}$ (blue circles). The black solid line is the sheet conductivity of the MoS$_2$ as a function of $V_g$. The insets show schematically the spin current path (green arrow) in the OFF state (left inset) and the ON state (right inset) of MoS$_2$.

The operation of the graphene/MoS$_2$ spin field-effect transistor relies on the absorption of spins traveling through the graphene by the MoS$_2$, as schematically illustrated in the inset of Fig. 3b. In order to support this argument, we make use of the spin resistances of the channel (graphene) and the absorbing material (MoS$_2$), which are the main control parameters in the spin absorption mechanism. In a broad approximation, they can quantify how easily the spin current flows through each of the materials, in the same way in which one can estimate a charge current flow in parallel electrical resistors. The spin resistances of graphene and MoS$_2$ can be expressed as $R_{Gr}^S = \frac{R_{Gr}^\blacksquare \lambda_{Gr}}{w_{Gr}}$ and $R_{MoS2}^S = \frac{R_{MoS_2}^\blacksquare (\lambda_{MoS_2})^2}{w_{Gr} w_{MoS_2}}$, respectively; being $R_{Gr,MoS_2}^\blacksquare = 1/\sigma_{Gr,MoS_2}^\blacksquare$ their sheet resistances, $\lambda_{Gr,MoS_2}$ their spin diffusion lengths and $w_{Gr,MoS_2}$ their widths. We have estimated the intrinsic spin lifetime in bulk MoS$_2$ to be in the range of 10 ps (see Supplementary Information). We have considered electron interaction with flexural phonons finding weak temperature dependence of the spin relaxation in accord with our experimental results and in contrast to the findings in ML MoS$_2$[13]. The lack of space inversion symmetry in ML has two effects on the spin transport. The first one is to increase the amplitude of the spin-flip matrix element[12]. The second effect is to induce spin-splitting of the energy bands at the K point. While the former effect enhances spin relaxation, the latter one suppresses it when the spin splitting is large enough to exclude elastic scattering. In p-type ML TMDs, for example, the spin splitting in the valence band is of the order of hundreds meV and the overall spin lifetime is prolonged compared with bulk. In this view, the spin degeneracy of the energy bands in few-layer TMDs renders these materials ideal spin sinks when put in proximity to graphene. Using the estimated spin lifetime in bulk MoS$_2$, we calculate $\lambda_{MoS_2} \sim 20$ nm in the OFF state of the device at $V_g = 40$ V. In contrast, the spin diffusion length in graphene is much longer, being $\lambda_{Gr} \sim 1.2$ μm estimated from Hanle measurements on a reference device (see Supplementary Information). Substituting the spin diffusion lengths, the electrical properties and geometrical factors of graphene and MoS$_2$, we obtain $R_{Gr}^S = 408$ Ω and $R_{MoS2}^S = 2.7$ Ω (see Supplementary Information). The fact that $R_{MoS_2}^S \ll R_{Gr}^S$ demonstrates the capability of the MoS$_2$ to absorb spins from the graphene channel. This is further supported by the low graphene/MoS$_2$ barrier height at high positive $V_g$[26], making the interface resistance sufficiently low for efficient spin absorption.

The situation completely changes when the gate voltage $V_g$ is swept towards negative values. At $V_g = -30$ V, the MoS$_2$ conductivity $\sigma_{MoS_2}^\blacksquare$ decreases by more than six orders of magnitude from $V_g = 40$ V, which leads to a similar increase in $R_{MoS_2}^S$. Therefore, we have $R_{MoS_2}^S \gg R_{Gr}^S$, which leads to no spin absorption by MoS$_2$. The inverse correlation between the spin signal $\Delta R_{nl}$ and the MoS$_2$ conductivity $\sigma_{MoS_2}^\blacksquare$ can be clearly seen in Fig. 3b. This correlation supports the aforementioned argument, and discards other scenarios, such as spin dephasing in possible trap states at the graphene/MoS$_2$ interface. The fact that similar results to those in Fig. 3b are also observed at 200 K indicates that the effect barely changes with temperature, being this experimental fact incompatible with the exponential temperature dependence expected for capture and escape in trap states (see Supplementary Information).

The spin absorption mechanism can be further confirmed by computing the expected spin signal ratio, $\Delta R_{nl}^{abs}/\Delta R_{nl}$, which quantifies the relative amount of spins deviating from the graphene channel towards the MoS$_2$[27]:

$$\frac{\Delta R_{nl}^{abs}}{\Delta R_{nl}} = \frac{2R_{MoS_2}^S\{\sinh(L/\lambda_{Gr})+2Q_I e^{L/\lambda_{Gr}}+2Q_I^2 e^{L/\lambda_{Gr}}\}}{R_{Gr}^S\{\cosh(L/\lambda_{Gr})-1\}+2R_{MoS_2}^S\sinh(L/\lambda_{Gr})+2R_I^S\{e^{L/\lambda_{Gr}}(1+Q_I)(1+2Q_{MoS_2})-1\}}, \quad (1)$$

where $\Delta R_{nl}^{abs}$ and $\Delta R_{nl}$ are the spin signals with and without spin absorption by the MoS$_2$; $R_I^S = R_I/(1-P_I^2)$ is the spin resistance of the Co/TiO$_2$/graphene interface, $R_I$ being the interface resistance and $P_I$ the interface spin polarisation; and finally, $Q_{MoS_2} = \frac{R_{MoS_2}^S}{R_{Gr}^S}$ and $Q_I = \frac{R_I^S}{R_{Gr}^S}$. Substituting the known parameters into Eq. (1), one gets $\Delta R_{nl}^{abs}/\Delta R_{nl} \approx 0.017$ at $V_g$ = 40 V (see Supplementary Information). The very small value calculated for $\Delta R_{nl}^{abs}/\Delta R_{nl}$ predicts a strong spin absorption, which confirms this scenario to be responsible for the observed experimental results of Fig. 3b.

The seamless integration of two 2D layered materials with remarkably different spin-orbit coupling amplitudes leads to a breakthrough device capable of both transporting and electrically controlling a spin current. Our approach, combined with recent advances in chemical production of high quality graphene[28] and TMD[22], as well as homostructural[29] and heterostructural[30] tunnel barriers for spin injection, may well lead to applications in the information and communication technology (ICT) sector. Furthermore, the van der Waals heterostructure at the core of our experiments also opens the path for fundamental research of exotic transport properties predicted for TMDs[12, 14-15], in which electrical spin injection has so far been elusive.

**Methods:**

**Device fabrication**. Fabrication of monolayer graphene samples uses the mechanical exfoliation method initiated in reference[24]. We first exfoliate bulk graphitic crystals onto a Nitto tape (Nitto SPV 224P) and repeat the cleavage process between 3-5 times until thin flakes can be identified visually by eye. The Nitto tape with relatively thin flakes is pressed against a preheated Si substrate with 300 nm $SiO_2$. After peeling off the Nitto tape, the substrate is examined under an optical microscope and monolayer graphene is identified by well-established optical contrast. We then prepare the $MoS_2$/PDMS stamp following reference[25]. First, a $MoS_2$ crystal is exfoliated twice using the Nitto tape and transferred on to a piece of poly-dimethyl siloxane (PDMS) (Gelpak PF GEL film WF x4, 17 mil.). After identifying the desired few-layer $MoS_2$ flake using optical contrast, it is transferred on top of graphene after slowly removing the viscoelastic stamp.

The lateral spin valve is formed following a standard nanofabrication procedure including electron-beam lithography; metal deposition and metal lift off in acetone. 5 Å of Ti are deposited by electron-beam evaporation and left to oxidize in air for 0.5 hours before depositing 35 nm of Co using electron-beam evaporation.

**Electrical measurements**

The measurements are performed in a Physical Property Measurement System (PPMS) by Quantum Design, using a "DC reversal" technique with a Keithley 2182 nanovoltmeter and a 6221 current source. A current bias of 10 µA is used unless stated in the text. Gate voltage is applied using a Keithley model 2636.

**Acknowledgements** The work at CIC nanoGUNE was supported by the European Union 7th Framework Programme under the Marie Curie Actions (607904-13-SPINOGRAPH) and the European Research Council (257654-SPINTROS), by the Spanish MINECO under Project No. MAT2012-37638, and by the Basque Government under Project No. PC2015-1-01. The work at the University of Rochester was supported by the Department of Energy under Contract No. DE-SC0014349, National Science Foundation under Contract No. DMR-

**Author Contributions**: F. C. conceived the study. W. Y., O. T. and R. L. performed the experiments. W. Y, O. T, L. H. E and F. C. and analysed the data, discussed the experiments and wrote the manuscript; H. D. performed theoretical calculations on the spin relaxation properties of $MoS_2$. All the authors contributed to the scientific discussion and manuscript revision. L. E. H. and F. C. co-supervised the work.

**Author Information**: The authors declare no competing financial interests. Correspondence and requests for materials should be addressed to L. E. H. or F. C. (l.hueso@nanogune.eu; f.casanova@nanogune.eu).

SUPPLEMENTARY INFORMATION

# A two-dimensional spin field-effect transistor

Wenjing Yan, Oihana Txoperena, Roger Llopis, Hanan Dery, Luis E. Hueso and Fèlix Casanova

**I. Sample characterisation**

Figure S1 shows the geometry of the device, as measured by atomic force microscopy (AFM). The width of the graphene and MoS$_2$ flakes are $w_{Gr}$ ~3 µm and $w_{MoS_2}$ ~0.4 µm respectively, and distances between the electrodes in the reference LSV and the graphene/MoS$_2$ LSV are ~1 µm and ~1.8 µm, respectively (Fig. S1a). The thickness of the MoS$_2$ flake is $t_{MoS_2}$ ~7 nm (see Fig. S1b).

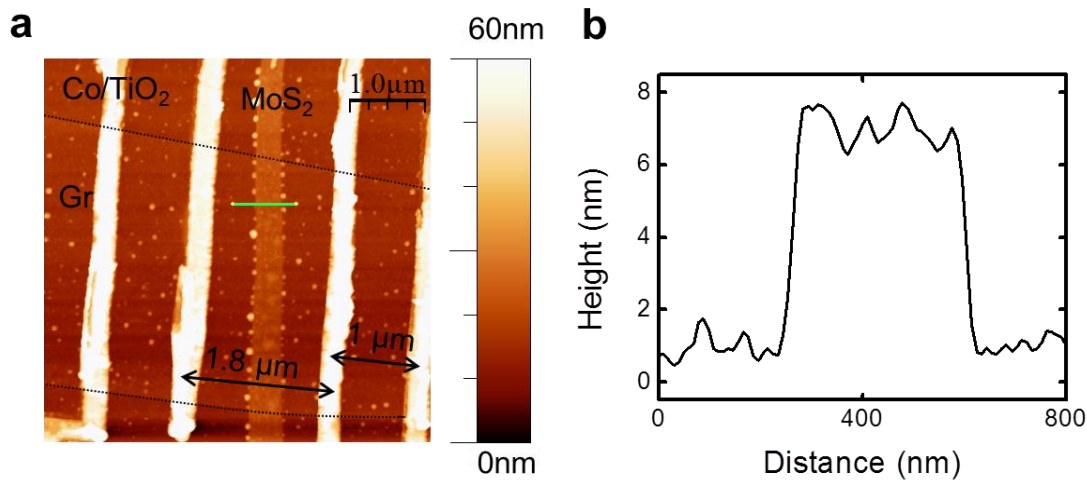

**Figure S1: AFM measurement on the devices. a**, Area scan showing the topography of the device, where graphene edge is traced by black dotted line and it is approximately 3 µm wide. The MoS$_2$ flake is intercepting the graphene flake. Distances between the Co/TiO$_2$ electrodes in the reference LSV and graphene/MoS$_2$ LSV are 1 µm and 1.8 µm, respectively. **b** Line profile taken from **(a)** across the MoS$_2$ flake on the marked line, where the thickness of the flake is extracted to be ~7 nm.

## II. Spin transport properties of graphene

The spin transport properties of graphene are studied in a typical graphene lateral spin valve as a function of temperature from 10 K to 300 K. The spin signal is weakly dependent on temperature, see Fig. S2a and references[8, 11].

The spin relaxation in the graphene channel was studied using Hanle precession. The experiment is done by first setting the injecting and detecting Co/TiO$_2$ electrodes in the parallel or antiparallel magnetisation state along the length of the electrodes by applying an in-plane magnetic field $B$. The device is then rotated by 90° and the non-local resistance $R_{nl}$ is measured while sweeping the magnetic field out-of-plane ($B_\perp$). Figure S2b shows the typical Hanle precession curve in the parallel (red circles) and antiparallel (blue circles) magnetization configurations of the electrodes, after removing the background signal arising from the out-of-plane tilting of the electrodes magnetization at high fields by subtracting the antiparallel from the parallel data[31].

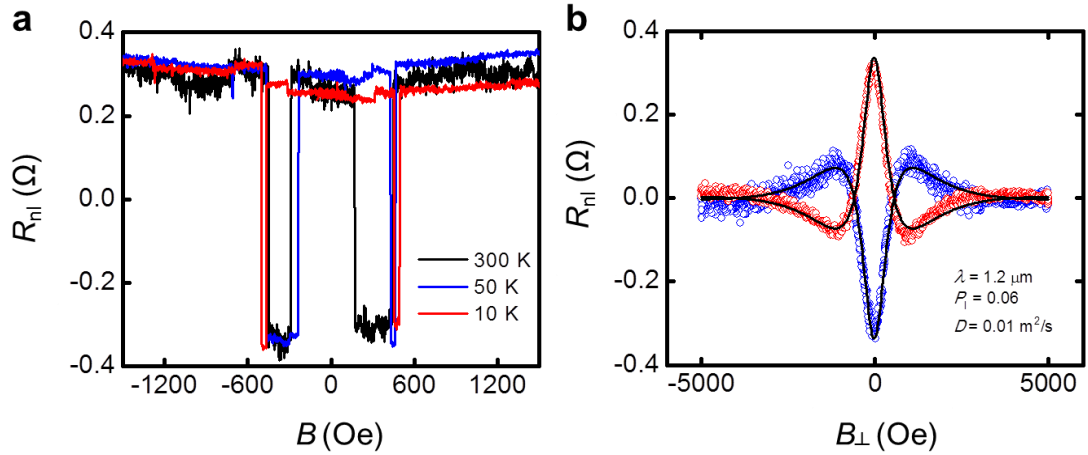

**Figure S2: Spin transport in a graphene lateral spin valve. a** Non-local resistance $R_{nl}$ measured with in-plane magnetic field sweep along the length of the electrodes at different temperatures. **b** Non-local resistance $R_{nl}$ is measured under out-of-plane magnetic field $B_\perp$ while the injecting and detecting Co/TiO$_2$ electrodes are parallel (red) and antiparallel (blue). The contribution from the out-of-plane rotation of the electrodes under $B_\perp$ is removed by subtraction the antiparallel curve from the parallel

curve. Spin diffusion length (λ), interface spin polarisation ($P_I$) and spin diffusion coefficient ($D$) are extracted by fitting Eq. S1 to the experimental data (black solid lines).

In order to fit the experimental data, we follow reference[32], which considers (i) spin precession and (ii) anisotropic spin absorption under the Co/TiO$_2$ injector contact. The following expression is used[32]:

$$R_{nl} = -2R_N \left( \frac{P_{F1}}{1-P_{F1}^2} \frac{R_{F1}}{R_N} + \frac{P_{I1}}{1-P_{I1}^2} \frac{R_{I1}}{R_N} \right) \left( \frac{P_{F2}}{1-P_{F2}^2} \frac{R_{F2}}{R_N} + \frac{P_{I2}}{1-P_{I2}^2} \frac{R_{I2}}{R_N} \right) \frac{C_{12}}{\det(\check{X})}, \quad (S1)$$

where $R_{Fk} = \rho_F \lambda_F / A_{Ik}$ are the spin resistances of the $k^{th}$ FM contact ($k$ =1 is the injector and $k$ = 2 is the detector), with resistivity $\rho_F$, spin diffusion length $\lambda_F$ and contact area of $A_{Ik}$; $R_N = \frac{R_{Gr}^\blacksquare \lambda_{Gr}}{w_{Gr}}$ is the spin resistance of graphene calculated with its sheet resistance ($R_{Gr}^\blacksquare$), its spin diffusion length ($\lambda_{Gr}$) and width ($w_{Gr}$); the resistance of the $k^{th}$ interface is $R_{Ik} = 1/G_{Ik}$, where $G_{Ik} = G_{Ik}^\uparrow + G_{Ik}^\downarrow$ is the effective conductance of the $k^{th}$ interface considering both spin up and down electrons, with $G_{Ik}^{\uparrow\downarrow} = 1/(2R_{Ik} + 2R_{Fk})$ ; $P_{Ik} = (G_{Ik}^\uparrow - G_{Ik}^\downarrow)/(G_{Ik}^\uparrow + G_{Ik}^\downarrow)$ describes the spin asymmetry (anisotropic spin absorption); and $C_{12}$ and $\det(\check{X})$ are defined as[32]:

$$C_{12} = -\det \begin{pmatrix} \text{Re}[\bar{\lambda}_\omega e^{-L/\bar{\lambda}_\omega}] & -\text{Im}[\bar{\lambda}_\omega e^{-L/\bar{\lambda}_\omega}] & -\text{Im}[\bar{\lambda}_\omega] \\ \text{Im}[\bar{\lambda}_\omega] & r_{1\perp} + \text{Re}[\bar{\lambda}_\omega] & \text{Re}[\bar{\lambda}_\omega e^{-L/\bar{\lambda}_\omega}] \\ \text{Im}[\bar{\lambda}_\omega e^{-L/\bar{\lambda}_\omega}] & \text{Re}[\bar{\lambda}_\omega e^{-L/\bar{\lambda}_\omega}] & r_{2\perp} + \text{Re}[\bar{\lambda}_\omega] \end{pmatrix}, \quad (S2)$$

$$\check{X} = \begin{pmatrix} r_{1\parallel} + \text{Re}[\bar{\lambda}_\omega] & \text{Re}[\bar{\lambda}_\omega e^{-L/\bar{\lambda}_\omega}] & -\text{Im}[\bar{\lambda}_\omega] & -\text{Im}[\bar{\lambda}_\omega e^{-L/\bar{\lambda}_\omega}] \\ \text{Re}[\bar{\lambda}_\omega e^{-L/\bar{\lambda}_\omega}] & r_{2\parallel} + \text{Re}[\bar{\lambda}_\omega] & -\text{Im}[\bar{\lambda}_\omega e^{-L/\bar{\lambda}_\omega}] & -\text{Im}[\bar{\lambda}_\omega] \\ \text{Im}[\bar{\lambda}_\omega] & \text{Im}[\bar{\lambda}_\omega e^{-L/\bar{\lambda}_\omega}] & r_{1\perp} + \text{Re}[\bar{\lambda}_\omega] & \text{Re}[\bar{\lambda}_\omega e^{-L/\bar{\lambda}_\omega}] \\ \text{Im}[\bar{\lambda}_\omega e^{-L/\bar{\lambda}_\omega}] & \text{Im}[\bar{\lambda}_\omega] & \text{Re}[\bar{\lambda}_\omega e^{-L/\bar{\lambda}_\omega}] & r_{2\perp} + \text{Re}[\bar{\lambda}_\omega] \end{pmatrix}, \quad (S3)$$

where $\bar{\lambda}_\omega = \tilde{\lambda}_\omega / \lambda_N$ with $\tilde{\lambda}_\omega = \lambda_N / \sqrt{i + i\omega_L \tau_{sf}}$ and the Larmor frequency $\omega_L = \gamma_e B_\perp = \frac{g\mu_B}{\hbar} B_\perp$; $r_{k\parallel} = \left( \frac{2}{1-P_{Ik}^2} \frac{R_{Ik}}{R_N} + \frac{2}{1-P_{Fk}^2} \frac{R_{Fk}}{R_N} \right)$; and $r_{k\perp} = \frac{1}{R_N G_{Ik}^{\uparrow\downarrow}}$.

For the fitting in Fig. S2b, we assume the injecting and detecting electrodes have (i) the same spin polarisation and (ii) the same interface resistances with the graphene channel. We fix the following experimental parameters: $P_{F1} = P_{F2} = 0.12$ (ref.[33]), $R_{I1} = R_{I2} = 10000\ \Omega$, $L = 2.26$ μm, $w_{Gr} = 0.73$ μm, $w_{F1} = 340$ nm, $w_{F2} = 230$ nm, $R_{Gr}^{\blacksquare} = 1317\ \Omega$, $\rho_F = 19\ \mu\Omega\text{cm}$ (ref.[33]), $\lambda_F = 40$ nm (ref.[34, 35]), $G_{Ik}^{\uparrow\downarrow} = 5 \times 10^{-5}\ \Omega^{-1}$, and obtain $P_I = 0.06$, $D = 0.01$ m$^2$/s, $\lambda_{Gr} = 1.2$ μm.

### III. Spin transport properties of MoS$_2$

We have made the following analysis and estimated the intrinsic spin relaxation time in bulk MoS$_2$ ($\tau_{MoS_2}$) is between ~10 ps and ~30 ps at 50 K. This was calculated via interaction of electrons with flexural phonons, which are long wavelength out-of-plane undulations. These phonons are far more populated than in-plane acoustic phonons (sound waves) since the interlayer van der Waals (vdW) interactions render the out-of-plane long wavelength undulations to be nearly "resistance-free" compared with in-plane motion of atoms (since atoms are held by strong chemical bonds in the plane). As important, the flexural phonons are strongly coupled to spin-flips[12]. It is also important to note that the analysis ignores extrinsic spin relaxation due to interaction with impurities and, therefore, if the MoS$_2$ is impurity-rich the $\tau_{MoS_2}$ value can be somewhat smaller. In addition, we ignore intervalley spin-flip scattering between K and K' valleys due to time-reversal symmetry (which applies to both monolayer and bulk MoS$_2$)[12].

Due to the fact that vdW interactions lead to weak interlayer coupling, the scattering is essentially a two-dimensional problem; that is, the electron motion before and after the scattering are mainly in-plane. Next, we assume that flexural phonons obey a quadratic dispersion law as often found in unstrained vdW materials[36, 37]. Specifically, $E_{ph} = \sqrt{\kappa \rho_m} q^2$, where $E_{ph}$ is the flexural phonon energy, $\kappa$ is the bending rigidity (~10 eV in MoS$_2$)[38], $\rho_m$ is the area mass density (~3·10⁻⁷ gr/cm² in MoS$_2$)[39, 40], and $q$ is the phonon wavevector.

Next, we use symmetry arguments to estimate the spin-flip matrix element of electrons due to scattering with flexural phonons in bulk MoS$_2$. Due to space inversion, the wavevector dependence of the spin-flip matrix element is quadratic, $M_{\text{bulk}} = D_{\text{so}} q^2$, where $D_{\text{so}}$ is scattering constant (units of energy·cm) coming from the spin-orbit coupling part of the deformation potential. In monolayers, where space inversion symmetry is not respected, the spin-flip matrix element follows a linear relation with the phonon wavevector, $M_{\text{mono}} = E_{\text{so}} q$, where $E_{\text{so}}$ is a spin-orbit coupling deformation potential, which for monolayer MoS$_2$ is $E_{\text{so}} \sim 0.2$ eV[41]. Given the spin-orbit coupling is non-vanishing only in the vicinity of the atomic cores, $E_{\text{so}}$ (monolayer) and $D_{\text{so}}$ (bulk) are related by $D_{\text{so}} \sim a\, E_{\text{so}}$, where $a$ is the lattice constant (~3 Å). Therefore, $D_{\text{so}} \sim 0.6$ eV·Å.

Using the dispersion of phonons and the spin-flip matrix element, the electron-phonon interaction that leads to intrinsic spin relation in the bulk is:

$$|< k,\uparrow |H_{\text{electron-phonon}}|k+q,\downarrow>|^2 = \left[\frac{k_B T}{2A\rho_m(\kappa/\rho_m)q^4}\right][D_{\text{so}}q^2]^2 = \frac{k_B T(D_{\text{so}})^2}{2A\kappa} \quad (S4)$$

where $k_B T$ is the thermal energy, being $k_B$ the Boltzmann constant and $T$ the temperature, and $A$ is the area of the flake. This area is cancelled when we turn the sum over all possible final scattered states into integral form. The interaction amplitude in the expression above is wavevector-independent due to cancelling effects between the phonon dispersion and spin-flip matrix element. This makes the summation over final states trivial, where the Fermi golden rule yields the following spin relaxation rate:

$$\frac{1}{\tau_{\text{MoS}_2}} = \frac{mk_\text{B}T(D_\text{so})^2}{\hbar^3 \kappa}, \tag{S5}$$

where $m$ is the electron mass. In the case of bulk $MoS_2$, $m$ is approximately the free electron mass. Plugging the numbers above, we get $\tau_{\text{MoS}_2} \sim 30$ ps. As shown in the expression, the spin relaxation rate dependence on T is linear, being much weaker than the exponential dependence found in monolayers[13]. The reason is that the bands are spin-degenerate in bulk $MoS_2$. In monolayers, on the other hand, the bands are spin-split and since spin-flips are largely elastic, the top spin-split band should be populated to have a non-negligible spin-flip amplitude.

Next, we consider that the dispersion of flexural phonons is renormalized due to the coupling between bending and stretching degrees of freedom[36]. This coupling prevents violent undulations and crumpling. In this case, the dispersion of flexural phonons follows is $E_\text{ph} = \sqrt[4]{k_\text{B}T/\rho_\text{m}}\sqrt{v_0}q^{3/2}$, where $v_0$ is the effective sound velocity ($5 \cdot 10^5$ cm/s). Repeating the analysis above, a renormalized electron-phonon interaction that leads to spin flips in the bulk is

$$\left| < k, \uparrow \left| H_\text{renorm.electron-phonon} \right| k+q, \downarrow > \right|^2 = \frac{q\sqrt{k_BT/\rho_m}(D_\text{so})^2}{2Av_0}. \tag{S6}$$

Since close to thermal equilibrium $q \sim \sqrt{2mk_\text{B}T}/\hbar$, the summation over final states in the Fermi golden rule yields the following spin relaxation rate

$$\frac{1}{\tau_{\text{MoS}_2}} = 4\pi \frac{mk_BT(D_\text{so})^2}{\hbar^4 v_0} \sqrt{\frac{2m}{\rho_m}}. \tag{S7}$$

Therefore, the relaxation remains linear in $T$. Plugging numbers we get $\tau_{\text{MoS}_2} = 10$ ps.

Next, we convert $\tau_{MoS_2}$ into $\lambda_{MoS_2}$ by using the diffusion coefficient $D_{MoS_2}$ as $\lambda_{MoS_2} = \sqrt{D_{MoS_2}\tau_{MoS_2}}$. $D_{MoS_2}$ depends on the mobility of the charge carriers in the $MoS_2$, $\mu_{MoS_2}$, as $D_{MoS_2} = \frac{\mu_{MoS_2} k_B T}{e}$.

In order to calculate $\mu_{MoS_2}$, we use the $\sigma^{\blacksquare}_{MoS_2}(V_g)$ data shown in Fig. 3b of the main text, which is representative of a typical $MoS_2$ flake transferred with the viscoelastic PDMS stamping (see methods) on a $SiO_2$ substrate. The results are reproducible from sample to sample because this transfer technique minimises the residues between the $MoS_2$ and the $SiO_2$ substrate and therefore enables an effective and repeatable field effect.

The $\sigma^{\blacksquare}_{MoS_2}(V_g)$ data in Fig. 3b has been measured for a fixed source-drain voltage of $V_{SD} = 0.26$ V and sweeping the gate voltage $V_g$ while measuring the current passing between source and drain, $I_{SD}$. From this data, we can calculate the mobility $\mu_{MoS_2}$ in the linear regime ($V_g \gtrsim -15$ V)[42] : $\mu_{MoS_2} = 84.6$ cm²/(V·s). Therefore, $D_{MoS_2} = 3.6 \cdot 10^{-5}$ m²/s.

Finally, we obtain $\lambda_{MoS_2} \approx 20$ nm. This value is necessary for calculating the spin resistance of the $MoS_2$, which is an important parameter for determining spin transport. Considering $\lambda_{MoS_2}$ is of the order of the thickness of $MoS_2$ (20 nm and 7 nm, respectively), one can get a more accurate estimation for $R^S_{MoS_2}$ by taking into account the length scale of the thickness of $MoS_2$ and $\lambda_{MoS_2}$ using the following equation[43]:

$$R^S_{MoS_2} = \frac{\rho_{MoS_2} \lambda_{MoS_2}}{w_{Gr} w_{MoS_2} \tanh(t_{MoS_2}/\lambda_{MoS_2})}, \tag{S8}$$

where $\rho_{MoS_2} = t_{MoS_2}/\sigma^{\blacksquare}_{MoS_2} = 6.4 \cdot 10^{-5}$ Ω·m is the resistivity of $MoS_2$. The hyperbolic tangent term comes from the boundary condition where the spin current $I_S = 0$ at the substrate, as detailed in Ref. 43. Eq. S8 results in $R^S_{MoS_2} = 2.8$ Ω. If we consider $t_{MoS_2} \ll \lambda_{MoS_2}$, Eq. S8 is simplified to:

$$R_{MoS_2}^S \approx \frac{R_{MoS2}^\blacksquare (\lambda_{MoS_2})^2}{w_{Gr} w_{MoS_2}}, \quad (S9)$$

where $R_{MoS_2}^\blacksquare = 1/\sigma_{MoS_2}^\blacksquare$ is the sheet resistance of MoS$_2$ which results in $R_{MoS_2}^S \approx 2.7\ \Omega$, very similar to the value obtained from Eq. S8. We consider the simplified expression of Eq. S9 for the analysis in this work.

**IV. Spin signal ratio calculation**

This section explains the details about the calculation for the spin signal ratio $\Delta R_{nl}^{abs}/\Delta R_{nl}$, being $\Delta R_{nl}^{abs}(\Delta R_{nl})$ the signal with (without) spin absorption by the MoS$_2$. We recall Eq. (1) from the main text for convenience[43]:

$$\frac{\Delta R_{nl}^{abs}}{\Delta R_{nl}} = \frac{2R_{MoS_2}^S \{\sinh(L/\lambda_{Gr}) + 2Q_I e^{L/\lambda_{Gr}} + 2Q_I^2 e^{L/\lambda_{Gr}}\}}{R_{Gr}^S \{\cosh(L/\lambda_{Gr}) - 1\} + 2R_{MoS_2}^S \sinh(L/\lambda_{Gr}) + 2R_I^S \{e^{L/\lambda_{Gr}}(1+Q_I)(1+2Q_{MoS_2}) - 1\}}, \quad (S10)$$

with $Q_{MoS_2} = R_{MoS_2}^S/R_{Gr}^S$ and $Q_I = R_I^S/R_{Gr}^S$, being $R_I^S = R_I/(1-P_I^2)$ the spin resistance of the Co/TiO$_2$/graphene contact and $P_I$ its spin polarization. Using all the parameters mentioned in the previous sections, we calculate $\Delta R_{nl}^{abs}/\Delta R_{nl} \approx 0.017$, predicting that the spin current traveling through the graphene channel is almost fully shunted by the MoS$_2$ flake in the on state of MoS$_2$. While Eq. S10 gives good enough qualitative understanding of the mechanism, it is assuming an approximate/simplified picture. A recent publication by Laczkowski *et al.* found the width of the absorber material is of paramount importance for the validity of Eq. (S10)[44]. The authors observe that, when this width becomes comparable to $\lambda$ of the material where the spins propagate, Eq. (S10) is not accurate anymore. This is due to the fact that Eq. (S10) describes spin absorption through a point-like contact between the two materials, and therefore ignores the spin accumulation profile under the spin absorber. The authors account for this by considering an effective $\lambda$ in the spin absorption area. Doing so, they calculate $\Delta R_{nl}^{abs}/\Delta R_{nl}$ values smaller than those obtained by Eq. (S10).

In our particular case, $w_{MoS_2}$ is comparable to $\lambda_{Gr}$, and therefore the correction proposed by Laczkowski *et al.* should be taken into account. This leads us to $\Delta R_{nl}^{abs}/\Delta R_{nl} < 0.007$, which further reinforces the fact that the MoS$_2$ acts as an extremely efficient spin absorber.

## V. Spin field-effect transistor

The spin field-effect transistor effect is robust and reproducible upon multiple gate sweeps and temperature cycles. Figure S3 shows non-local resistance measurements at different gate voltages performed before the ones shown in the main text (Fig. 2 and 3) in a different cryostat. Furthermore, to rule out charging effects as the origin of the spin transistor effect when sweeping the gate voltage, the measurements were performed in a random order: -30 V, 50 V, 0 V, 20 V, -10 V and 10V. Same to that observed in Fig. 3 of the main text, at large positive gate voltage the spin transport channel is completely turned OFF and no spin signal is measured, confirming the robustness and reproducibility of the effect.

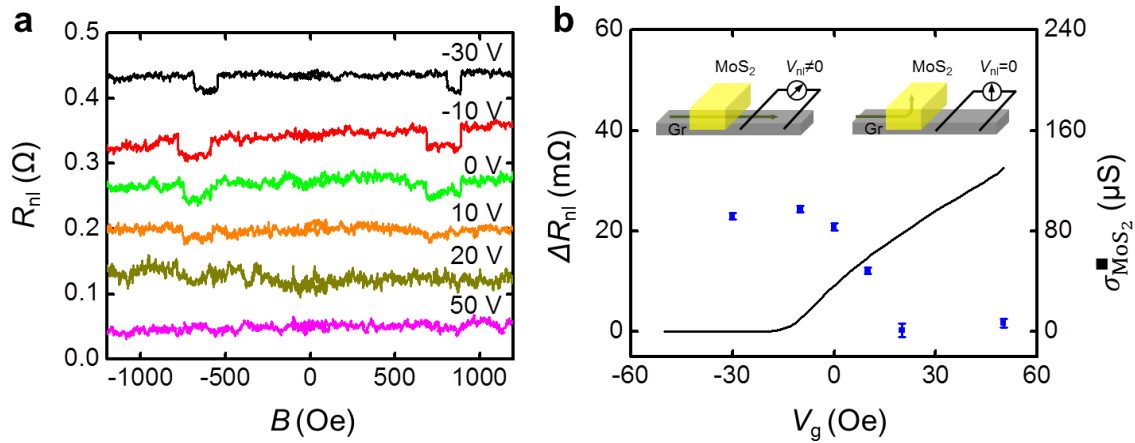

**Figure S3. Spin transport in the graphene/MoS$_2$ lateral spin valve at 50 K. a**, Non-local resistance $R_{nl}$ measured at different $V_g$ at 50 K using 10 µA current bias and for a centre-to-centre distance between ferromagnetic electrodes (*L*) of 1.8 µm. Individual sweeps are offset in $R_{nl}$ for clarity. **b**, Gate modulation of the spin signal $\Delta R_{nl}$ (blue circles). The black solid line is the sheet conductivity of the MoS$_2$ as a function of $V_g$. The insets show schematically the spin current path (green arrow) in the off state (left inset) and the on state (right inset) of MoS$_2$. The measurements were performed in a different cryostat than the one used in the main text and taken in the following order of gate voltage: -30 V, 50 V, 0 V, 20 V, -10 V, 10 V, confirming the robustness and reproducibility of the effect.

Figure S4 shows the spin transport measurements at 200 K for the devices shown in the main text. The spin signal of the reference graphene LSV at 200 K is similar to that at 50 K (compare Fig. S4a and Fig. 2a in the main text), as expected from the results in Fig. S2a and the literature[8,11], thus providing a fundamental building block on which room temperature spin field-effect transistor can be built.

The comparison between Fig. S4a and Fig. S4b demonstrates the operation of the spin field-effect transistor at 200 K. While $\Delta \boldsymbol{R_{nl}}$ in the reference graphene LSV varies slightly with $V_g$ (Fig. S4a), gate modulation of spin signal between ON and OFF states is clearly observed in the graphene/MoS$_2$ LSV (Fig. S4b), evidencing that the spin absorption of MoS$_2$ does not depend on the temperature. This experimental fact rules out the scenario of spin dephasing in trap states at the graphene/MoS$_2$ interface, because it is incompatible with the exponential temperature dependence expected for capture and escape in trap states, and suggests a weak temperature dependence of the spin relaxation time of few-layer MoS$_2$, as predicted by our calculations (see section III above).

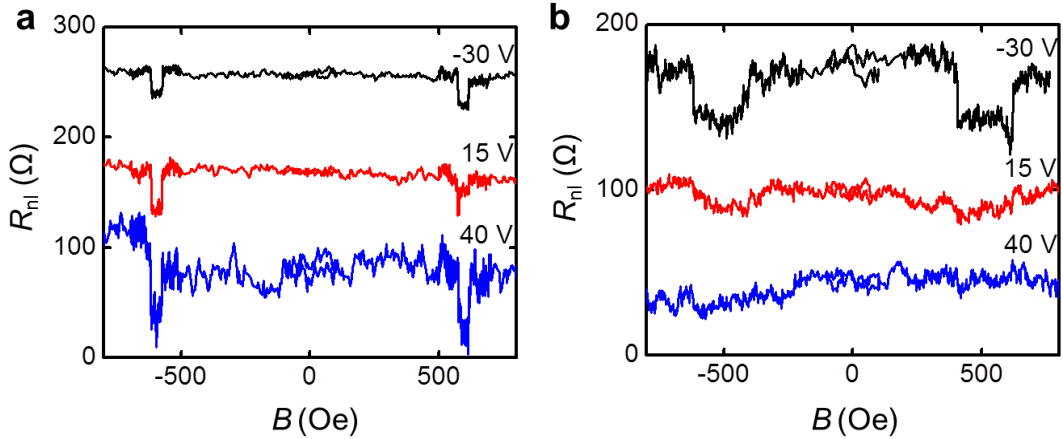

**Figure 4. Spin transport in reference graphene and graphene/MoS$_2$ lateral spin valves at 200 K.** Non-local resistance $R_{nl}$ measured at different $V_g$ using 10 µA current bias and for (**a**) the reference graphene LSV, in which the centre-to-centre distance between ferromagnetic electrodes (*L*) is 1 µm. (**b**) the graphene/MoS$_2$ LSV, in which the centre-to-centre distance between ferromagnetic electrodes (*L*) is 1.8 µm. Individual sweeps are offset in $R_{nl}$ for clarity.

**Additional References:**